\RequirePackage{fix-cm}
\documentclass[smallextended]{svjour3}       % onecolumn (second format)
\smartqed  % flush right qed marks, e.g. at end of proof
\usepackage{graphicx}
\usepackage{subfigure}
\begin{document}

\title{Do quantum strategies always win?%\thanks{Grants or other notes
%about the article that should go on the front page should be
%placed here. General acknowledgments should be placed at the end of the article.}
}
%\subtitle{Do quantum strategies always win? \\ If so, write it here}

\titlerunning{Do quantum strategies always win?}        % if too long for running head

\author{Namit Anand         \and
        Colin Benjamin %etc.
}
%\authorrunning{Short form of author list} % if too long for running head

\institute{N. Anand \at
              National Institute of Science Education and Research, Bhubaneswar - 751005, India 
             %\email{fauthor@example.com}           %  \\
%             \emph{Present address:} of F. Author  %  if needed
           \and
          Colin Benjamin \at
           National Institute of Science Education and Research, Bhubaneswar - 751005, India \\   
Tel.: +91-674-2304056\\
              Fax: +91-674-2304070\\
  \email{cbiop@yahoo.com}       }
\date{Received: date / Accepted: date}
% The correct dates will be entered by the editor

\maketitle

\begin{abstract}
 In a seminal paper, Meyer [David Meyer, Phys. Rev. Lett. 82, 1052 (1999)] described the advantages of quantum game theory by looking at the classical penny flip game. A player using a quantum strategy can win against a classical player almost 100\% of the time. Here we make a slight modification to the quantum game, with the two players sharing an entangled state to begin with. We then analyze two different scenarios, first in which quantum player makes unitary transformations to his qubit while the classical player uses a pure strategy of either flipping or not flipping the state of his qubit. In this case the quantum player always wins against the classical player. In the second scenario we have the quantum player making similar unitary transformations while the classical player makes use of a mixed strategy wherein he either flips or not with some probability ``p''. We show that in the second scenario, 100\% win record of a quantum player is drastically reduced and for a particular probability ``p'' the classical player can even win against the quantum player. This is of possible relevance to the field of quantum computation as we show  that in this quantum game of preserving versus destroying entanglement a particular classical  algorithm can beat the quantum algorithm.
% \subclass{MSC code1 \and MSC code2 \and more}
\end{abstract}

\section{Introduction}
\label{intro}
Game theory is an extremely interesting and sophisticated  field which holds within its ambit the power to resolve conflicts, propose new strategies in making war and peace and  {understand various situations wherein competing entities interact. Examples range from zero sum games like tick-tack-toe to non zero sum games like Prisoners dilemma.} However, in the past 15 years a new addition to game theory has come into being. This is the story of Quantum games. It studies how to quantize the previously played classical games or propose new games which explicitly use quantum mechanical phenomena like entanglement, nonlocality and quantum interference. Further, in contradistinction to the just mentioned definition of quantum games, sometimes a player may use quantum mechanics as a viable strategy to defeat his/her opponent in a classical game which is also included in the general framework of quantum games. 
 In the context of ``quantized games'', Meyer was one of the first to detect and prove that Quantum strategies if used by a player in the classical Penny flip game can help a player outwit her opponent 100\% to nil.  This led to a flourishing field of quantum games. Previous works include quantum prisoners dilemma \cite{ewl}, including an experimental implementation  of the Prisoner's dilemma in a NMR quantum computer\cite{Du}.  One of the key motivations for playing games, in the quantum world, comes from the possibility of re-formulating quantum communication protocols, and algorithms, in terms of games between quantum and classical players\cite{az}. In fact Ref.\cite{az} claims that there exists a basic relationship between quantum algorithms and quantum games and research on quantum games can lead to new quantum algorithms\cite{guo}.  

However there have been works which have commented that quantum games do not offer anything new and whatever they promise can be replicated via classical correlated equilibrium\cite{levine}. In this work we apply this hypothesis to the entangleD quantum penny flip game.  We see that indeed a classical player can outwit her quantum opponent by using a particular mixed strategy which is not possible in the non-entanglement based quantum penny flip game. 
 Further, Ref. \cite{enk}, questions two aspects of quantum game theory- first,  whether the quantum strategy solves the underlying classical problem and to what extent the new solution can be obtained via  a classical model. Ref. \cite{enk} gives the example of Shor's algorithm, in the language of game theory, this would be called a quantum strategy which solves the classical  factoring problem (the game).  Ref. \cite{enk}  also questions that the quantum prisoners dilemma doesn't really solve the classical prisoners dilemma. In the entangled quantum penny flip game of ours we are looking at a new game of preserving  versus destroying a  maximally entangled state in which counter intuitively a classical mixed strategy can defeat the hitherto unbeatable quantum strategy of Meyer's penny flip game\cite{Meyer}.  In other words, making an analogy to Shor's algorithm (quantum strategy) which solves the classical factoring problem (classical game), our work shows that a classical mixed strategy (classical algorithm) can beat the quantum strategy (quantum algorithm) in winning (solving) the entangled penny flip game (quantum problem). Stretching this analogy further, this therefore could in fact be of relevance to the field of quantum algorithms whose justification stems from the fact that they are indeed more efficient than classical algorithms. We in this work put forth a counter example which demonstrates that a particular classical  algorithm can outwit the previously unbeatable quantum algorithm in the entangled quantum penny flip problem.
 
 In recent years a move to  distinguish between the application of quantum mechanics to game theory known as ``quantized games'' and the introduction of game theory to quantum mechanics known as ``gaming the quantum''  has also been made. Reference \cite{simon} delineates this distinction in proper detail.``Gaming the quantum'' refers to the fact that gamed quantum systems should collapse to the underlying classical game on introduction of restrictions that would imply that one has effectively quantized the game. However, in the absence of any restrictions one is gaming the quantum\cite{simon}. So if a gamed quantum system does not reduce to the underlying classical game on application of certain restrictions then it is an example of a quantized game while one which reduces to the underlying classical game on imposing those restrictions is an example of gaming the quantum. Gaming the quantum is acknowledged as the better approach to understanding quantum game theory than just quantizing the game\cite{simon}. In this work we show how our ``entangled penny flip game'' under certain restrictions(no entanglement and no superposition)  collapses to the classical matching pennies game, thus is an example of ``gaming the quantum'' rather than of just ``quantizing the game''. All gaming the quantum examples are also examples of quantized games however this isnt true in the reverse. Quantized games like EWL's quantum prisoners dilemma\cite{ewl} are those which do not reduce to their underlying classical game on imposing restrictions.
 
The entangled quantum penny flip game could be mistaken for the two penny flip game\cite{bala}. In the latter there are two pennies to begin with, however in the former there is a Bell state to begin with, which is the original penny of the classical penny flip. The two states of the penny, heads and tails are now the maximally entangled or the completely separable states. There is however a crucial difference between the classical penny flip and the entangled quantum penny flip. In the former there is no `draw'  while in the latter the final state could be in an non-maximally entangled state, in which case the game ends in a draw.  

 Our aim is to provide an example from the field of quantum games wherein  veritably the first time a classical algorithm or strategy can beat a quantum algorithm or strategy. Generally, it is almost lazily assumed that a quantum algorithm will be better than a classical algorithm. We challenge this assumption through this example.

In the sections below we first introduce the classical matching pennies and an alteration to it- the Penny flip game. Next, we explain the Quantum version\cite{Meyer},  we then introduce the quantum entangled  penny flip game and then analyze cases wherein the classical player uses a pure strategy and finally  wherein the classical player uses a mixed strategy. This is followed by a section on the question of ``quantized games'' versus ``Gaming the quantum'' wherein we show our example game falls in the category of gaming the quantum- the right way to approach quantum games. Finally, we give a brief perspective on future endeavors, quantum circuit implementation and conclusions at the end.
\section{ Classical Penny flip game} PQ Penny flip game as discussed by Meyer in his paper\cite{Meyer} "Quantum strategies" is a altered version of the traditional classical game ``Matching pennies'' which has the following rules:
\subsection{Matching pennies}
\begin{itemize}
\item Players P and Q each have a penny and are separated from each other so they cant communicate, however there is an unbiased Referee with whom they communicate simultaneously and disclose the state of their pennies.  
\item Initial state of the pennies is either of (H,H), (H,T), (T,H) or (T,T).
\item Each player can choose to either flip or not flip his/her own penny, independent of each other.
\item If at the end of the game both the pennies are heads or both are tails, then P wins else Q wins. 
\end{itemize}

Thus, the payoff matrix of the game is:
\begin{equation}%
\begin{array}
[c]{c}%
{P}%
\end{array}%
\begin{array}
[c]{c}%
Head\\
Tail
\end{array}
\stackrel{%
\begin{array}
[c]{c}%
{Q}%
\end{array}
}{\stackrel{%
\begin{array}
[c]{cc}%
Head\hskip .5cm &  Tail
\end{array}
}{\left[
\begin{array}
[c]{cc}%
(1,-1) & (-1,1) \\
(-1,1) & (1,-1)
\end{array}
\right]  }} \label{DiffractionPD}%
\end{equation}
The numbers in the matrix above are the payoffs for either player, first index is for P and second is for Q.  For example $(-1,1)$ means P loses a penny while Q gains a penny as end state is Heads.
This is a zero sum game and there is no pure strategy Nash equilibria here. However there is a mixed strategy Nash equilibria\cite{azhar}. In the mixed strategy the players repeatedly play the game and as has been shown before in Ref.\cite{azhar} P and Q both getting exactly 50 \% of the times Heads and Tails is the mixed strategy Nash equilibrium at which the payoff of both players is zero.  
\subsection{PQ Penny flip}

The PQ penny flip game was designed by Meyer in Ref.\cite{Meyer}, its a close cousin of the Matching pennies game as described above and has the following rules:
\begin{itemize}
\item Players P and Q each have access to a single penny. 
\item Initial state of the penny is heads(say).
\item Each player can choose to either flip or not flip the penny and if in the end the state is heads, Q wins else P wins.
\item Players cannot see the current state of the penny.
\item Sequence of actions : Q $\longrightarrow$  P $\longrightarrow$  Q 
\item If final state is heads, Q wins else P wins
\end{itemize}

Since players cannot see the current state of the penny, their actions become independent of each other and so each strategy, flipping or not flipping is equally desirable.\\
The matrix form of the game is:
\begin{equation}%
\begin{array}
[c]{c}%
{P}%
\end{array}%
\begin{array}
[c]{c}%
N\\
F
\end{array}
\stackrel{%
\begin{array}
[c]{c}%
{Q}%
\end{array}
}{\stackrel{%
\begin{array}
[c]{cccc}%
NN\hskip .5cm  &  NF\hskip .5cm  &  FN \hskip .5cm &  FF
\end{array}
}{\left[
\begin{array}
[c]{cccc}%
(-1,1) & (1,-1) & (1,-1) & (-1,1)\\
(1,-1) & (-1,1) & (-1,1) &(1,-1)
\end{array}
\right]  }} \label{DiffractionPD}%
\end{equation}
where N is not flipping and F is flipping. Both players have $\frac{1}{2}$ as their winning probability. The numbers in the matrix above are the payoffs for either player, first index is for P and second is for Q.  For example $(-1,1)$ means P loses a penny while Q gains a penny as end state is Heads. PQ penny flip is an example of a strictly competitive or zero sum game which again has no pure strategy Nash equilibrium but has as before a mixed strategy Nash equilibrium.  The pair of mixed strategies of P flipping the penny with probability $\frac{1}{2}$  and Q playing each of the available four strategies with probability $\frac{1}{4}$ is the  mixed strategy Nash equilibrium with payoff again zero.

\section{ Quantum Penny Flip game} The Quantum penny flip game as described by Meyer in Ref.\cite{Meyer} is as follows. The starship Enterprise
is facing a calamity. This is when Q appears on the bridge and offers to
rescue the ship if Captain P can beat him at a simple game: Q produces a penny and asks the captain to place it in a small box,
head up. Then Q, followed by P, followed by Q, dip their fingers into the box,
without looking at the penny, and either flip it over or leave it as it is.
After Q's second turn they open the box and Q wins if the penny is head up.
 Q wins every time they play, using the following ``quantum'' strategy:

\begin{eqnarray*}
  \left|  0\right\rangle &\stackrel {\mbox{\small Q does $H$ }}{\longmapsto}&
\frac{1}{\sqrt{2}}(\left|  0\right\rangle +\left|  1\right\rangle )\\
&  \stackrel{\mbox {\small P does $X$ or $I$}}{\longmapsto}&\frac{1}{\sqrt{2}}(\left|  0\right\rangle +\left|
1\right\rangle )\\
&  \stackrel{\mbox{\small  Q does $H$}}{\longmapsto}&\left|  0\right\rangle
\end{eqnarray*}
Here $0$ denotes `head' and $1$ denotes `tail', ${H}=\frac{1}{\sqrt{2}%
}\left(
\begin{array}
[c]{cc}%
1 & 1\\
1 & -1
\end{array}
\right)  $ is the Hadamard transformation , $I$ means
leaving the penny alone and the action with $X=\left(
\begin{array}
[c]{cc}%
0 & 1\\
1 & 0
\end{array}
\right)  $ flips the penny over. Q's quantum strategy of putting the penny
into the equal superposition of `head' and `tail', on his first turn, means
that whether Picard flips the penny over or not, it remains in an equal
superposition which Q can rotate back to `head' by applying ${H}$ again
since ${H}={H}^{-1}$. So Q always wins when they open the box. Thus playing the penny flip game with a quantum strategy enables the player to win against one who is playing classical, 100\% of the time. Of course one may ask what if both play quantum? In that case as shown by Meyer, the quantum advantage vanishes, see Theorem 2 of Ref.\cite{Meyer}:A two-person zero-sum game need not have a quantum/quantum equilibrium.

\section{ Quantum entangled penny flip game: Introduction} 
The game that we introduce here has the following interpretation, we begin with a maximally entangled state of two qubits which is shared by P and Q and allow P and Q to make moves on only the qubit in their possession. If the final state of the game is a maximally entangled state then Q wins, and if it is a non maximally entangled state then its a draw, if its a  separable state then P wins. We allow P to make classical moves, i.e. either $I$ or $X$ and Q to make Hadamard transforms (or any unitary operations) on his  qubit. This game should not be confused with the quantum two penny game\cite{bala} wherein there are two heads and two tails, here the heads of the penny is the ``entangled state'' and tails is the ``separable state''. It is the entanglement analog of the classical penny flip game with the addition  of a ``draw''.

\subsection{Quantum entangled penny flip game: The classical pure strategy}
In this case the classical player P is allowed only the pure strategy of either flipping or not flipping his qubits.
Consider the initial state of the system as the entangled state: 
\begin{equation}
\left|  \psi \right\rangle  = \frac{1}{\sqrt{2}} (\left|  10\right\rangle - \left|  01\right \rangle)
\end{equation}
Sequence of actions is : Q $\longrightarrow$  P $\longrightarrow$  Q.

{\it Step 1}:

So, if Q does a Hadamard transformation on his qubit we get:
\begin{equation}
H \otimes I (\frac{1}{\sqrt{2}} (\left|  10\right\rangle - \left|  01\right\rangle)) = \frac{1}{2} (\left|  00\right\rangle - \left|  01\right\rangle - \left|  10\right\rangle - \left|  11\right\rangle)
\end{equation}

So the current state of the game is:
\begin{equation}
 \left|  \psi \right\rangle = \frac{1}{2} (\left|  00\right\rangle - \left|  01\right\rangle - \left|  10\right\rangle - \left|  11\right\rangle)
\end{equation}

{\it Step 2}:

Now its P's turn to make a move, which means either leaving the state unchanged or applying the flip $X$ operation, when applying the flip the state becomes-
\begin{equation}
I \otimes X (\frac{1}{2} (\left|  00\right\rangle - \left|  01\right\rangle - \left|  10\right\rangle - \left|  11\right\rangle)) = \frac{1}{2} (\left|  01\right\rangle - \left|  00\right\rangle - \left|  11\right\rangle - \left|  10\right\rangle)
\end{equation}

So the current state of the game is either of:
\begin{equation}
 \left|  \psi \right\rangle = \frac{1}{2} (\left|  01\right\rangle - \left|  00\right\rangle - \left|  11\right\rangle - \left|  10\right\rangle) 
\end{equation}
or
\begin{equation}
 \left|  \psi \right\rangle = \frac{1}{2} (\left|  00\right\rangle - \left|  01\right\rangle - \left|  10\right\rangle -\left|  11\right\rangle) 
\end{equation}

{\it Step 3}:

Q's final move on the above states leaves us with:
\begin{equation}
H \otimes I (\frac{1}{\sqrt{2}} \left|  01\right\rangle - \left|  00\right\rangle - \left|  11\right\rangle - \left|  10\right\rangle ) = \left|  B1\right\rangle  = \frac{1}{\sqrt{2}} (\left|  11\right\rangle - \left|  00\right\rangle)
\end{equation}
or 
\begin{equation}
H \otimes I ( \frac{1}{2} (\left|  00\right\rangle - \left|  01\right\rangle - \left|  10\right\rangle - \left|  11\right\rangle)) = \frac{1}{\sqrt{2}} (\left|  00\right\rangle + \left|  11\right\rangle) 
\end{equation}

So finally we have the either the  following states, depending on whether $P$ had flipped or not flipped his qubit:
\begin{itemize}
\item $\left|  B\right\rangle  = \frac{1}{\sqrt{2}} (\left|  11\right\rangle - \left|  00\right\rangle) $
\item $\left|  B1\right\rangle  = \frac{1}{\sqrt{2}} (\left|  11\right\rangle + \left|  00\right\rangle)$
\end{itemize}

Matching pennies with entangled states and one player having quantum strategies while the other is classical, like the original version by Meyer gives a definite win to Q as both the states above are maximally entangled Bell states.
The game here is about whether player Q having all quantum strategies at his hand can keep the state maximally entangled, whereas P with classical moves can or cannot reduce the entanglement. If in the end all the states obtained are maximally entangled then Q wins, if they are separable then P wins. What the game essentially shows is that it is not possible for player P with pure classical strategies to effect the quantum correlations between the particles in a way which Q cannot revive with her quantum moves.

\subsection{ Quantum entangled penny flip game : The  classical mixed strategy}  Now we allow for P using albeit classical but mixed strategy. This entails P with  probability $p$ flipping the state of his qubit and with probability $1-p$ leaving it as it is. In this case as before P and Q start will a maximally entangled Bell pair. Q makes the first move, again a Hadamard.  Next is P's turn and as defined earlier flips with probability $p$. Finally Q does a Hadamard. At the end the final state is observed for the amount of entanglement.  As before P and Q share a Bell pair. But contrary to the previous case, P now uses a mixed classical strategy, Q still uses the pure quantum strategy. Since P uses a mixed strategy we have to take recourse to density matrices to explain the results. 

{\it Step 1: The initial state }

In the form of density matrices the initial state $ \left| \psi  \right\rangle = \frac{1}{\sqrt{2}}[\left|  10 \right\rangle -\left|  01 \right\rangle]$ is represented as-

$\rho_{0}= \left|  \psi \right\rangle \left \langle  \psi \right| =\frac{1}{2}\left[\begin{array}
[c]{cccc}%
0 & 0 & 0 &0\\
0 & 1& -1 &0\\
0&-1&1&0\\
0&0&0&0
\end{array}\right]$.

{\it Step 2: Q makes her move}

Q makes an unitary transformation on her part of the shared state. $U_{Q_{1}}=\left[\begin{array}
[c]{cc}%
a & b^{*} \\
b & -a^{*}\\
\end{array}\right]$. The state after Q's move then is $\rho_{1}=(U_{Q1}\otimes I)\rho_{0} {(U_{Q1}\otimes I)}^{\dagger}$.
\begin{figure}
\centering{\includegraphics[scale=0.5]{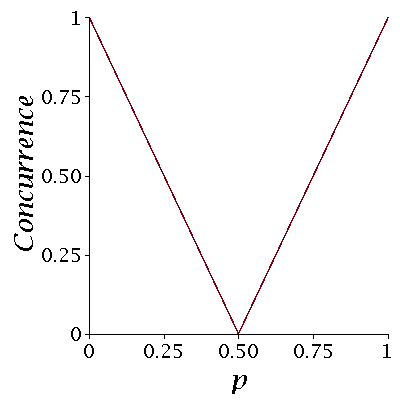}} \caption{  {Concurrence vs p showing that entanglement vanishes at $p = 1/2$, so by P's classical moves entanglement is completely destroyed enabling him to win.}} \label{neg} 
\end{figure}
 
{\it Step 3: P plays mixed}

P as we said earlier in contrast to the previous section plays a mixed strategy, which entails flipping the state of his qubit with probability ``p'' or not flipping.
The state after P's move then is:  $\rho_{2}=p( I \otimes X) \rho_{1} ( I \otimes X)^{\dagger} +(1- p)(I \otimes I) \rho_{1} (I \otimes I)^{\dagger}$.

{\it Step 4: Q makes her final move}

At the end Q makes her final move, which as before has to be an unitary transformation, it further could be same as her first move or different. Thus $U_{Q2}=\left[\begin{array}
[c]{cc}%
\alpha & \beta^{*} \\
\beta & -\alpha^{*}\\
\end{array}\right]$. The state after this final  move then is $\rho_{3}=(U_{Q2}\otimes I)\rho_{2}(U_{Q2}\otimes I)^{\dagger}$.

 To understand this case of P using mixed, lets analyse this case for $Q$ using the familiar Hadamard transform in both steps 2 and 4. 
 In this special case, \\
 \begin{center}
 $\rho_{3}=\frac{1}{2}\left[\begin{array}
[c]{cccc}%
p & 0 & 0 &-p\\
0 & 1-p& -1+p &0\\
0&-1+p&1-p&0\\
-p&0&0&p
\end{array}\right]$. 
\end{center}

To check the entanglement content of this final state we take recourse to an entanglement measure- Concurrence . Concurrence for a two qubit density matrix $\rho_3$ is defined as follows- we first define a "spin-flipped" density matrix, $\gamma$ as $(\sigma_y \otimes \sigma_y) \rho_{3}^* (\sigma_y \otimes \sigma_y) $. Then we calculate the square root of the eignevalues of the matrix $\rho_{3} \gamma$  (say $\lambda_1$,$\lambda_2$,$\lambda_3$,$\lambda_4$) in decreasing order. Then, Concurrence is :\begin{center}
max ($\lambda_1-\lambda_2-\lambda_3-\lambda_4,0$)
\end{center}
In Fig. 1,  the Concurrence is plotted. At $p=1/2$ concurrence vanishes implying a separable state and a win for P's mixed classical strategy.   Thus proving our contention that there exists a classical mixed strategy which can defeat a quantum strategy.

\subsubsection{Quantum entangled penny flip game: General quantum strategy versus mixed classical strategy}
In the above two cases we had the Quantum player playing the identical quantum strategy of applying a Hadamard at his turn. Now what if he uses a general unitary not restricted to just a Hadamard. Further in successive turns he does not implement the same unitary, i.e., $U_{Q_{1}}\neq U_{Q_{2}}$. We again see that as in the previous case when confronted with P playing a mixed classical strategy the quantum player is beaten for the case when $p=1/2$.
We implement general quantum strategy by just two changes to the scheme introduced in the above sub-section. Replacing the Hadamard in Step 2 with $U_{Q_{1}}$ and  the Hadamard in Step 4 with $U_{Q_{2}}$. $U_{Q_{i}}=\left[\begin{array}
[c]{cc}%
\cos(\theta_{i})e^{i\phi_{i}} & \sin(\theta_{i})e^{i\phi_{i}^{\prime}} \\
 \sin(\theta_{i})e^{-i\phi_{i}^{\prime}}&-\cos(\theta_{i})e^{-i\phi_{i}}  \\
\end{array}\right], i=1,2$.

In Fig. 2 we plot the Concurrence when the quantum player does not play with just Hadamard but does a general unitary transformation to the qubit in his possession.  We only plot the Concurrence versus $\theta_1$ and $\theta_2$, to show that the classical player wins for $p=1/2$. One can similarly plot the Concurrence versus $\phi_{1},\phi_{2}, \phi_{1}^{\prime},\phi_{2}^{\prime}$ and get identical results to what has been plotted in Fig. 2, so we do not repeat them here.
\begin{figure*}
\centering
\subfigure[Concurrence vs. $\theta_{1}$, $\theta_{2}=0,\phi_{1}=\pi/2, \phi_{1}^{\prime}=0, \phi_{2}=\pi/2, \phi_{2}^{\prime}=0$]{\includegraphics[width=.49\textwidth]{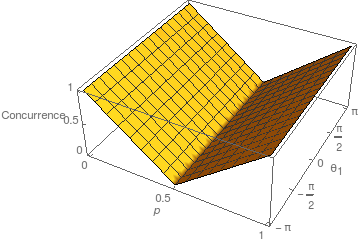}}
\subfigure[Concurrence vs. $\theta_{2}$, $\theta_{1}=0,\phi_{1}=\pi/2, \phi_{1}^{\prime}=0, \phi_{2}=\pi/2, \phi_{2}^{\prime}=0$]{\includegraphics[width=.49\textwidth]{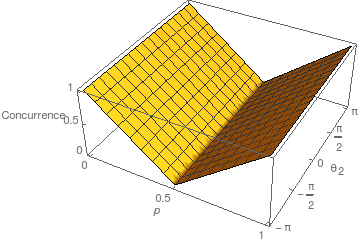}}
 \caption{The Concurrence when quantum player plays a general unitary vs. classical players mixed strategy. The classical player always wins when $p=1/2$, confirming that regardless of whether quantum player uses a Hadamard or any other unitary he always loses when classical player plays a mixed strategy of either flipping or not flipping with probability 50\%.} 
\end{figure*}

\section{Quantum circuit implementation}
\begin{figure*}
\centering
\subfigure[Quantum player uses  Hadamard.]{\includegraphics[width=.49\textwidth]{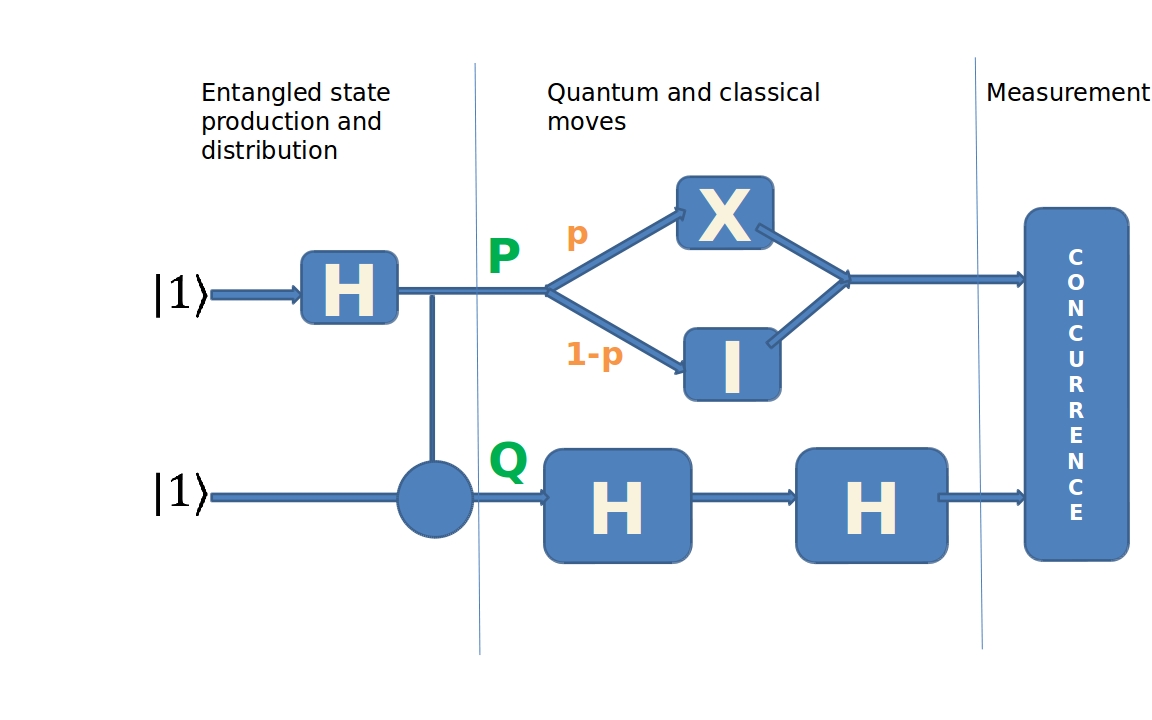}}
\subfigure[Quantum player uses a general unitary]{\includegraphics[width=.49\textwidth]{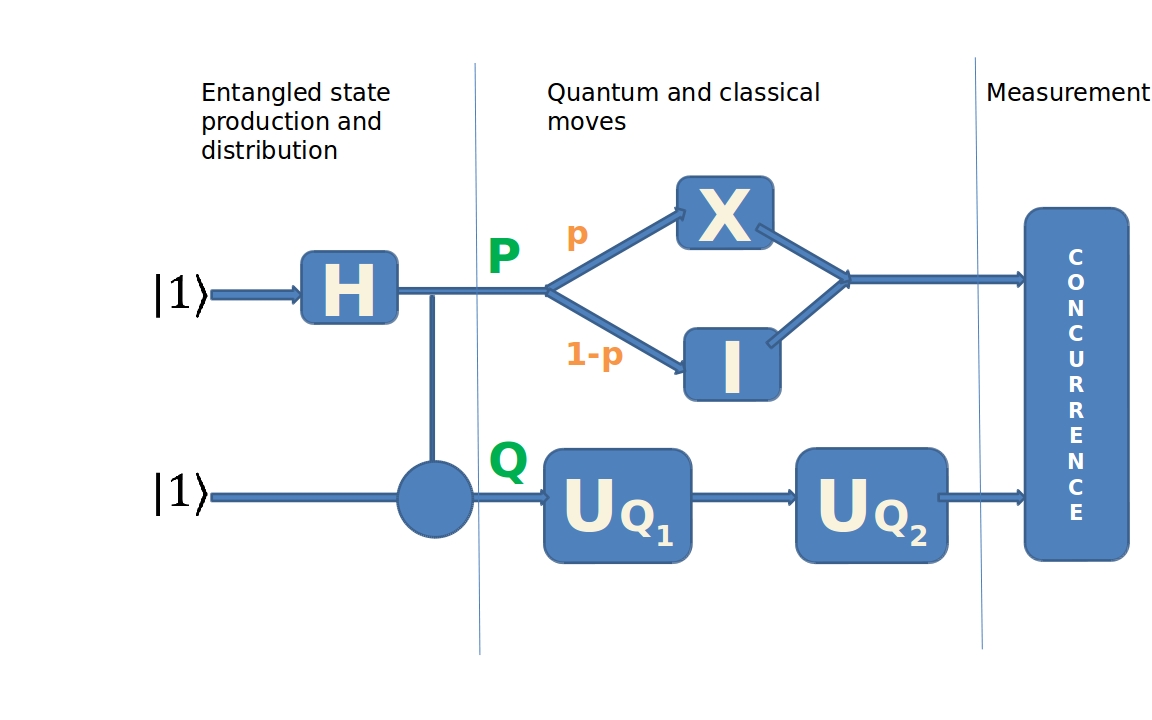}}
\caption{The quantum circuit for the entangled penny flip game. M denotes measurement of entanglement content via concurrence, see Ref.\cite{walborn}.} 
\end{figure*}

To design the quantum circuit for this game, we first have a Referee/Arbiter who generates the Bell state $\frac{1}{\sqrt{2}}(\left|01 \right\rangle-\left|10 \right\rangle)$ as shown in Fig. 3. He then hands over this state to P and Q, who now share this maximally entangled 2-qubit state. {P and Q get access to their respective qubits from the Referee. The operations they undertake on their respective qubits proceeds as $Q\rightarrow P\rightarrow Q$. Q first does a unitary (Hadamard) on her qubit and then P with probability p flips his qubit or with probability 1-p leaves it unaltered. Finally, Q does another unitary (Hadamard, again). The Referee now comes back into the picture where he takes the final output state and establishes the entanglement content via the Concurrence. The Concurrence of a general  two qubit mixed state has been experimentally determined in Ref.\cite{walborn}. If the Referee finds that the Concurrence vanishes (Separable state) then P is declared winner, on the other hand if the Concurrence is one (Maximal entanglement), then Q wins, in the rest of the cases the honors are equally shared i.e. it's a draw.}

\section{Gaming the quantum vs. quantized games: Finding the underlying classical game}
In Ref. \cite{simon} the authors delineate the ideas in the theory of quantum games into introducing elements of game theory in quantum mechanics(gaming the quantum) vs using quantum mechanics to study game theory (quantized games). Our entangled  quantum penny flip game is an example of gaming the quantum, since when restrictions are put ``no entanglement'' and ``no superposition'' the game reduces to the classical matching pennies game.

Meyer in his paper Quantum strategies uses the PQ penny flip game closely related to the traditional Matching pennies game. The following restrictions transform our Quantum entangled penny flip game to the original Matching pennies game. 
If we destroy the entanglement between the qubits and allow both the players to make only classical moves (Q has no recourse now to a Hadamard and has to satisfy herself with a flip or no-flip), the results show the same outcomes as the matching pennies game as shown below: 

 We begin with the referee destroying the entangled state
$
\left|  \psi \right\rangle  = \frac{1}{\sqrt{2}} (\left|  10\right\rangle - \left|  01\right \rangle)
$  shared between P and Q:
 The state of the system is now either $\left|  \psi \right\rangle  = \left|  10\right\rangle $ or $ \left|  01\right\rangle $ .
The sequence of actions by either player is same as before: Q $\longrightarrow$  P $\longrightarrow$  Q,  which puts the state of the system in either of the following four product states: 
$\left|  \psi \right\rangle  = \left|  00\right\rangle $ or $ \left|  01\right\rangle $ or $\left|  10\right\rangle $ or $ \left|  11\right\rangle $.

 If the state is $\left|  \psi \right\rangle  = \left|  00\right\rangle $ or $ \left|  11\right\rangle $, i.e., both bits in the same state then Q wins.
 and if the state is $\left|  \psi \right\rangle  = \left|  01\right\rangle $ or $ \left|  10\right\rangle $ i.e., both bit states dont match then P wins.
In this way our Quantum entangled penny flip game reduces to the underlying Matching pennies game. Hence satisfying the criteria for ``gaming the quantum'' as described in ref \cite{simon}.

\section{Conclusions}
The Quantum penny flip game with entangled particles has  nontrivial outcomes as compared to the original quantum penny flip game\cite{Meyer}. In a particular case where classical player uses a mixed strategy with p = ``0.5'', the quantum player indeed loses as opposed to the expected win for all possible unitaries!\\ The first takeaway from this work with particular relevance to game theory  is that some one using a classical strategy can beat someone with a quantum strategy. Meyer showed that in the PQ penny flip if both players use Quantum strategies then there is no advantage.  However, a player using a quantum strategy will win 100\% of the time against a player using a classical strategy because of an enlarged strategy space due to quantum superpositions. We wanted to check the generality of this result. Since the claim is usually made that quantum strategies are more powerful than classical strategies, we find that this result does not hold in our case of entangled quantum penny flip game. So the moral of the story is that quantum strategies are not(always) better than classical strategies. 

 The second takeaway is of relevance to the wider world of quantum computation wherein quantum algorithms have been shown to be more efficient than classical algorithms, for example Shor's algorithm. We in this work put forth a counter example which demonstrates that a particular classical algorithm can outwit the previously unbeatable quantum algorithm in the entangled quantum penny flip problem. On top of that the mixed strategy works against any possible unitary as we show by simulation on a strategy space for all possible parameters.

Finally, we show that our game is an example of ``gaming the quantum'' rather than just a quantized game which anyway is a special case of gaming the quantum. Our future endeavors include extending such a mixed strategy to a more genral class of games, checking for the effect of noise and phenomena like decoherence in entanglement based games. 

\begin{acknowledgements} Colin Benjamin, would like to thank Dept. of Science and Technology (Nanomission), Govt. of India, for funding via Grant No. SR/NM/NS-1101/2011.
\end{acknowledgements}

\end{document}